# Josephson junction in cobalt-doped BaFe$_2$As$_2$ epitaxial thin films on (La, Sr)(Al, Ta)O$_3$ bicrystal substrates

Takayoshi Katase[a*], Yoshihiro Ishimaru[b], Akira Tsukamoto[b], Hidenori Hiramatsu[c], Toshio Kamiya[a], Keiichi Tanabe[b] and Hideo Hosono[a,c] a: Materials and Structures Laboratory, Mailbox R3-1, Tokyo Institute of Technology, 4259 Nagatsuta-cho, Midori-ku, Yokohama 226-8503, Japan b: Superconductivity Research Laboratory, International Superconductivity Technology Center (ISTEC), 10-13 Shinonome 1-chome, Koto-ku, Tokyo 135-0062, Japan c: Frontier Research Center, S2-6F East, Mailbox S2-13, Tokyo Institute of Technology, 4259 Nagatsuta-cho, Midori-ku, Yokohama 226-8503, Japan

**ABSTRACT**

Josephson junctions were fabricated in epitaxial films of cobalt-doped $BaFe_2As_2$ on [001]-tilt $(La,Sr)(Al,Ta)O_3$ bicrystal substrates. 10-µm-wide microbridges spanning a 30-degrees-tilted bicrystal grain boundary (BGB bridge) exhibited resistively-shunted-junction (RSJ)-like current–voltage characteristics up to 17 K, and the critical current was suppressed remarkably by a magnetic field. Microbridges without a BGB did not show the RSJ-like behavior, and their critical current densities were 20 times larger than those of BGB bridges, confirming BGB bridges display a Josephson effect originating from weakly-linked BGB.

--------------------------------------------------------

Footnotes:

[*] Electronic mail: katase@lucid.msl.titech.ac.jp

The discovery of iron pnictide superconductors in early 2008[1] rekindled the extensive material research on superconductors.[2-4] Materials in this new system possess attractive properties, including high critical temperatures ($T_c$) up to 56 K[5] and high critical magnetic fields (upper critical fields > 100 T).[6,7] These properties have aroused active research on epitaxial films and device applications of new high-$T_c$ superconductors. Recently, we successfully fabricated epitaxial films of LaFeAsO[8] and superconducting epitaxial films of cobalt-doped $AE$Fe$_2$As$_2$ ($AE$ = Sr and Ba) with $T_c$ of ~20 K.[9,10] Additionally, several groups have reported the fabrication of epitaxial films of Fe-based superconductors such as fluorine-doped $Ln$FeAsO ($Ln$ = La, Nd),[11-14] cobalt-doped $AE$Fe$_2$As$_2$,[15-19] and FeSe$_{1-x}$Te$_x$[20] with high $T_c \geq 20$ K. However, the crystal qualities, superconducting properties, and chemical stabilities must typically be improved to prepare superconducting devices. We initially tried to create a Josephson junction using cobalt-doped SrFe$_2$As$_2$ (SrFe$_2$As$_2$:Co) epitaxial films, but were unsuccessful due to the insufficient critical current density ($J_c$)[21] and high reactivity even with water vapor in air.[22] However, we later found that cobalt-doped BaFe$_2$As$_2$ (BaFe$_2$As$_2$:Co) epitaxial films are stable even in a moist atmosphere.[10] In addition, BaFe$_2$As$_2$:Co films exhibit an improved film quality compared to SrFe$_2$As$_2$:Co; e.g., BaFe$_2$As$_2$:Co films have a better crystallinity and atomically-flat surfaces, although some droplets and pits defects remain.

Lee *et al.*[15] fabricated bicrystal grain boundary (BGB) junctions in BaFe$_2$As$_2$:Co epitaxial films on [001]-tilt SrTiO$_3$ (STO) bicrystal substrates, and reported the dependence of $J_c$ on the misorientation angle of the BGB as well as the suppression of $J_c$ across a BGB. Moreover, they reported growth of BaFe$_2$As$_2$:Co epitaxial films and concluded that to obtain a high $J_c$ of ~5 MA/cm$^2$ (at 4.2 K, self-field), a buffer layer such as (001) STO and BaTiO$_3$ (BTO) is necessary for (La,Sr)(Al,Ta)O$_3$ (LSAT) substrates.[18] Although the Josephson effect using a single-crystal of Fe-based superconductors has been reported,[23-25] a Josephson junction based on epitaxial thin films, which is essential for practical device applications, has yet to be reported.

In this letter, we demonstrate a Josephson junction in an epitaxial film of a Fe-based superconductor. Distinct Josephson effects are observed in microbridges with BGB fabricated in BaFe$_2$As$_2$:Co epitaxial films grown on [001]-tilt LSAT bicrystal substrates.

BaFe$_2$As$_2$:Co epitaxial films (250 nm in thickness) were deposited on LSAT (001) single crystal and [001]-tilt LSAT bicrystal substrates with a 30-degree misorientation angle by pulsed laser deposition (PLD). The second harmonic of a Nd:YAG laser (wavelength: 532 nm) operating at a repetition rate of 10 Hz was used as the excitation source, and sintered disks of stoichiometric BaFe$_{1.84}$Co$_{0.16}$As$_2$ were employed as PLD targets. High quality epitaxial films were grown directly on LSAT (001) substrates without a buffer layer by the following process. In a previous paper,[10] BaFe$_2$As$_2$:Co epitaxial films contained several % of Fe impurities and showed a superconducting transition width of $\Delta T_c = 3$ K and $J_c \ll 1$ MA/cm$^2$. Probably due to insufficient film quality, we could not obtain an operating Josephson junction with these films. Therefore, we improved the growth process to reduce the Fe impurity and to increase film uniformity, in particular by improving the phase purity of the PLD targets and the homogeneity of the substrate temperature. The PLD targets were synthesized by solid-state reactions of a stoichiometric mixture of BaAs, Fe$_2$As, and Co$_2$As via the reaction of BaAs + 0.92Fe$_2$As + 0.08Co$_2$As → BaFe$_{1.84}$Co$_{0.16}$As$_2$. A key to obtaining a single-phase BaFe$_2$As$_2$:Co was to employ a finely-cut metal Ba reagent to complete the reaction of the starting materials. In this work, we employed growth conditions at 850 °C and in a vacuum of ~$10^{-5}$ Pa.

Figure 1(a) shows the out-of-plane x-ray diffraction (XRD, anode radiation: Cu K$\alpha_1$) pattern of the improved BaFe$_2$As$_2$:Co epitaxial film on a LSAT (001) substrate. Intense peaks of BaFe$_2$As$_2$:Co 00$l$ and LSAT 00$l$ diffractions were observed, but peaks due to a metal Fe impurity, which should appear at 2$\theta$ ~44 and 65 degrees,[10] were almost removed. Figure 1(b) shows the temperature dependence of the resistivity for the BaFe$_2$As$_2$:Co films. Similar to those of single-crystals[26] and recently-reported epitaxial films,[18,19] an onset transition temperature ($T_c^{onset}$) of ~22 K and a sharper transition temperature width ($\Delta T_c$) of ~1.1 K were obtained. The film showed a high $J_c$ of 2 – 10 MA/cm$^2$ at 3 K.

The BGB bridge structures were defined by photolithography and Ar ion milling. The BGB microbridge was 300-μm long and 10-μm wide. The $I$–$V$ characteristics were measured by the four-probe method without and under magnetic fields perpendicular to the film. Figure 2(a) shows the in-plane $\phi$ scan of the 200 diffraction of a BGB bridge chip. Eight diffraction peaks, which were classified into two groups with a 90-degree rotational periodicity, were observed, confirming that the two crystals with the four-fold rotational symmetry coexist with a tilt angle of 30 degrees. This result verifies that the BaFe$_2$As$_2$:Co BGB bridge has an epitaxial relationship with the LSAT bicrystal substrate. The resistivity measurements of the BGB bridge showed a sharp $\Delta T_c$ = 1.2 K and $T_c^{onset}$ = 22.6 K, both of which are the same as those obtained for the BaFe$_2$As$_2$:Co epitaxial film in Fig. 1(b).

Figure 2(b) shows the current ($I$) – voltage ($V$) chacteristics of the BGB bridge measured at 11 K, which exhibited a critical current ($I_c$) of 1.5 mA ($J_c$ of 60 kA/cm$^2$) without hysteresis. The shape of the $I – V$ curve indicates a resistively-shunted-junction (RSJ) type behavior at temperatures from 17 to 4.2 K. The estimated normal-state resistance $R_N$ of the BGB bridge was ~0.012 Ω, and the $R_N A$ product ($A$ is the cross-sectional area of the junction) was $3.0 \times 10^{-10}$ Ωcm$^2$, which is smaller by an order of magnitude than that of a typical Josephson junction of a YBa$_2$Cu$_3$O$_{7-\delta}$ (YBCO) epitaxial film on a STO bicrystal substrate.[27] This result contrasts the case of a YBCO Josephson junction; i.e., it is considered that the low carrier density regions formed due to facets and a distorted lattice would result in a large $R_N$ for the YBCO Josephson junction.[28] We tentatively attribute the small $R_N$ of the BaFe$_2$As$_2$:Co Josephson junction to the metallic nature of the normal state.

In contrast, the $I – V$ characterstics of microbridges without BGB (non-BGB bridge) did not show a RSJ-like behavior and only exhibited a resistivity jump due to the normal-state transition at $J_c$ of ~2 MA/cm$^2$ at 11 K. This result demonstrates that, even if the microbridge region contains domain/grain boundaries, a boundary with a small rotation mismatch does not work as a Josephson junction.

The uniformity of the current distribution across the junction was examined by measuring $I_c$ of the BGB bridge as a function of the magnetic field ($B$) perpendicular to the film. The $I – V$ curves at $B = 0$ and ~0.9 mT in Fig. 3(a) demonstrate that $I_c$ is clearly suppressed by applying a magnetic field at 10 K. The magnitude of $I_c$ modulation, which is defined as $[I_c(0) – I_c(0.9 \text{ mT})] / I_c(0)$, was ~95%, indicating that most of the suppercurrent though the BGB bridge is due to the Josephson current. Figure 3(b) shows the $I_c – B$ pattern at 10 K. (The yellow arrow indicates the sweep direction.) The $I_c – B$ pattern showed a hysteresis, and the maximum $I_c$ was observed at nonzero field values (e.g., at ~0.7 mT as indicated by the red triangle). The presence of hysteresis is attributed to the influence of the flux pinnig in the vicinity of the BGB, similar to the case of BGB Josephson junctions using YBCO epitaxial films.[29] The $I_c - B$ curve did not show a clear multiple oscillation pattern and deviated from the ideal Fraunhofer pattern expected for a small junction with a uniform current distribution. Additionally, short-period oscillations, which are similar to a superconducting quantum interference device and attributed to nonuniform current flow across the junction due to inhomogeneous barriers and the existence of high resistance regions such as defects and dislocations, were observed. Considering the London penetration depth ($\lambda_L$ ~300 nm) of BaFe$_2$As$_2$:Co at ~10 K,[30] the Josephson penetration depth $\lambda_J$ was calculated by the following fomula, $\lambda_J = (\hbar / 2e\mu_0 d J_c)^{1/2}$ where $d$ is the effective penetration depth $2\lambda_L$ ignoring barrier thicknesses, $\hbar$ plank constant divided by $2\pi$, $e$ the elementary electric charge, and $\mu_0$ the vacuum permeability. Using the observed $J_c$ value of 88 kA/cm$^2$ gave a $\lambda_J$ value of 1.8 μm, indicating that the junction is a "large junction" where the junction width (10 μm) is much larger than $\lambda_J$. This suggests that the Josephson current flows in the edge regions of the junction.

Figure 4(a) shows the temperature dependece of $J_c$ of the BGB bridge and a non-BGB bridge. The $J_c$ of the BGB bridge is ~20 times smaller than that of the non-BGB bridge throughout the entire temperature range. For the BGB bridge, the supperconducting current was observed up to 18 K. Figure 4(b) shows the temperature dependence of the $I_cR_N$ product of the BGB bridge, which demosntrates that the value decreased almost linearly with increasing temperature. The inset is the $R_N$–$T$ plot. $R_N$ almost remained unchanged at $T < 12$ K, which is the same behavior as the BGB Josephson junctions of YBCO epitaxial films.[27] $R_N$ of the present system increased gradually with increasing temperautre at higher $T$. This dependence differs from that of the YBCO case where the $R_N$ value is almost unchanged up to $T_c$.

In summary, we demonstrated Josephson junctions with 10 μm-wide microbridges across a bicrystal grain boundary in BaFe$_2$As$_2$:Co epitaxial films on LSAT [001]-tilt bicrystal substrates. The $I$ – $V$ curve displayed RSJ-like characteristics from 17 K to 4.2 K. The critical current was clearly modulated by an external magnetic field applied perpendicular to the film surface. We expect the present achievement will lead to the development of superconducting devices and circuits based on the epitaxial technology of Fe-based high-$T_c$ superconductors.

This work conducted at the International Superconductivity Technology Center (ISTEC) was supported by New Energy and Industrial Technology Development Organization (NEDO), Japan. The present work was supported in part by the Funding Program for World-Leading Innovative R&D on Science and Technology, Japan.

**Figure captions**

Fig. 1. (Color online) (a) out-of-plane XRD pattern at room temperature and (b) Temperature ($T$) dependence of electrical resistivity ($\rho$) of a BaFe$_2$As$_2$:Co epitaxial film.

Fig. 2. (Color online) (a) In-plane $\phi$ scan of 200 diffractions of the BGB bridge on the LSAT bicrystal substrate at room temperature. (b) $I - V$ characteristic of the BGB bridge at $T = 11$ K.

Fig. 3. (Color online) (a) $I - V$ curves of the BGB bridge under $B = 0$ and 0.9 mT at $T = 10$ K. (b) Magnetic field dependence of the critical current ($I_c$) of the BGB bridge at 10 K. Yellow arrow and red triangle indicate the direction of the magnetic field sweep and the maximum $I_c$ position, respectively.

Fig. 4. (Color online) (a) Temperature ($T$) dependences of critical current densities ($J_c$) of the microbridges with BGB ('BGB bridge') and without BGB ('non-BGB bridge'). (b) Critical current density – normal state resistance ($I_cR_N$) product as a function of temperature for the BGB bridge. Inset shows the temperature dependence of $R_N$.

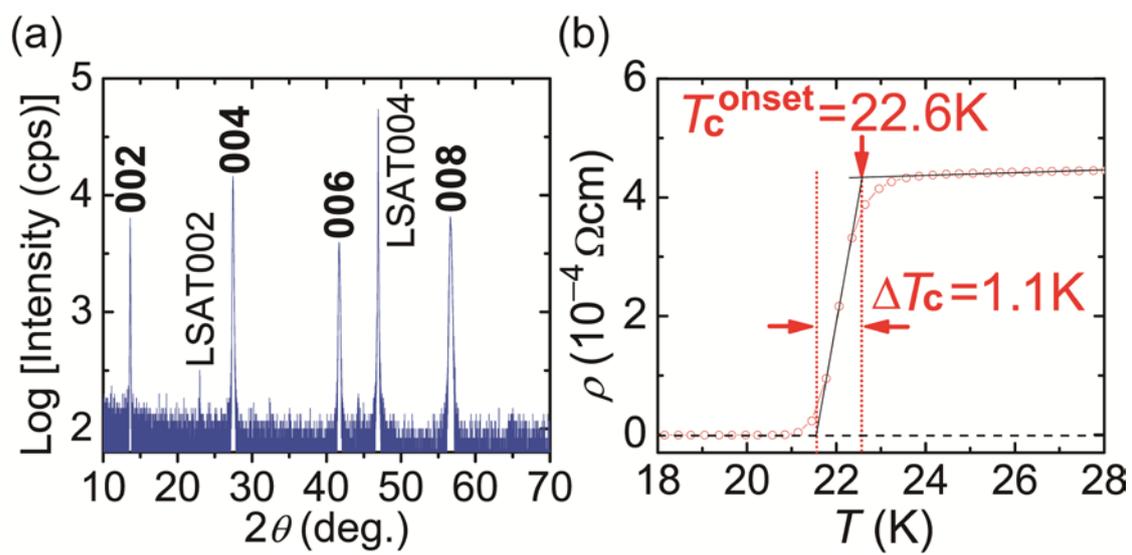

Fig. 1.

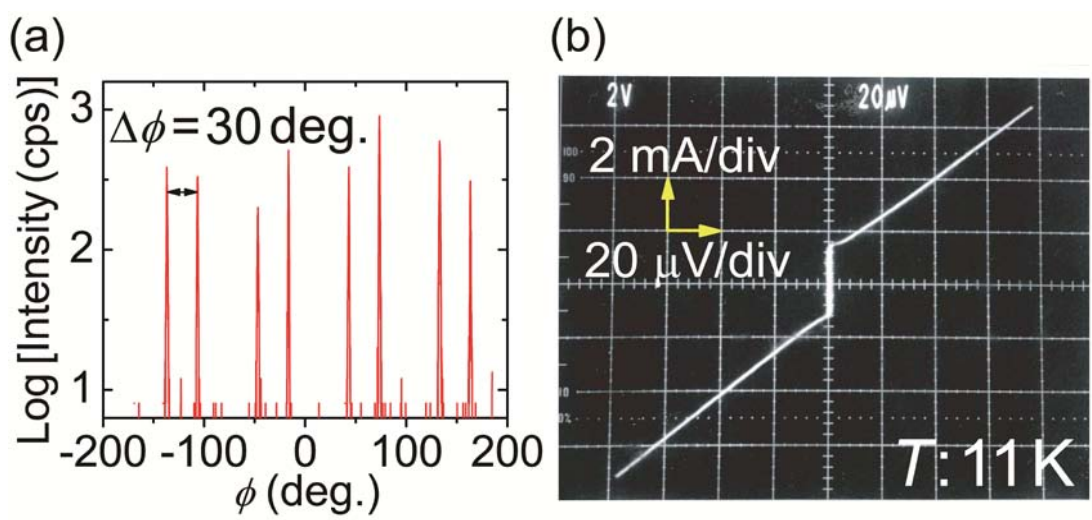

Fig. 2.

T.Katase et al.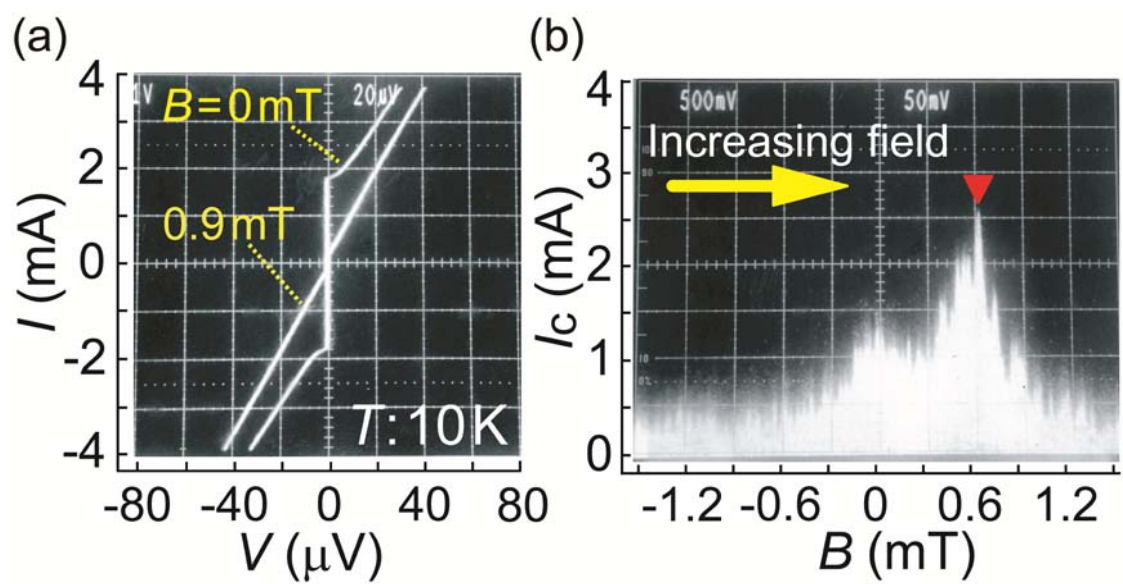

Fig. 3.

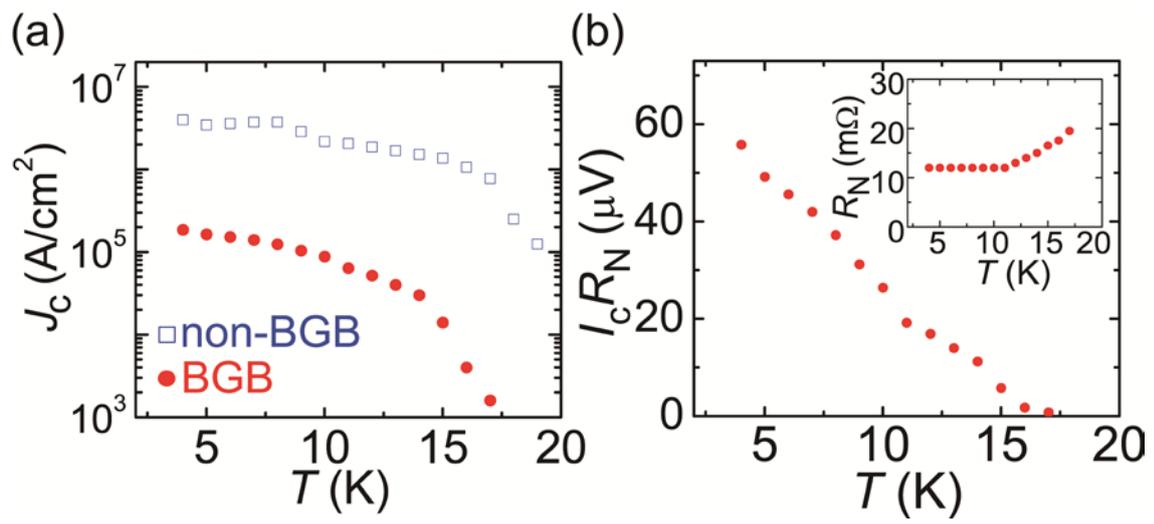

Fig. 4.